\documentclass{article}

\usepackage{cite,url,amssymb,amsmath,graphics,epsfig}
\usepackage{color}

\definecolor{code}{rgb}{0.8,0,0}
\definecolor{output}{rgb}{0,0,0}

\makeatletter
\newcommand{\contraction}[5][1ex]{%
  \mathchoice
    {\contraction@\displaystyle{#2}{#3}{#4}{#5}{#1}}%
    {\contraction@\textstyle{#2}{#3}{#4}{#5}{#1}}%
    {\contraction@\scriptstyle{#2}{#3}{#4}{#5}{#1}}%
    {\contraction@\scriptscriptstyle{#2}{#3}{#4}{#5}{#1}}}%
\newcommand{\contraction@}[6]{%
  \setbox0=\hbox{$#1#2$}%
  \setbox2=\hbox{$#1#3$}%
  \setbox4=\hbox{$#1#4$}%
  \setbox6=\hbox{$#1#5$}%
  \dimen0=\wd2%
  \advance\dimen0 by \wd6%
  \divide\dimen0 by 2%
  \advance\dimen0 by \wd4%
  \vbox{%
    \hbox to 0pt{%
      \kern \wd0%
      \kern 0.5\wd2%
      \contraction@@{\dimen0}{#6}%
      \hss}%
    \vskip 0.5ex%  how far above the line starts
    \vskip\ht2}}

\newcommand{\contraction@@}[3][0.05em]{%
  \hbox{%
    \vrule width #1 height 0pt depth #3%
    \vrule width #2 height 0pt depth #1%
    \vrule width #1 height 0pt depth #3%
    \relax}}
\makeatother

\begin{document}

\title{Computer Algebra calculation of XRMS polarisation dependence in the non spherical case}
\author{Thomas A. Wood\footnote{Van Mildert College, University of Durham, DH1 3LH, England}, Alessandro MIRONE\\
ESRF, BP 220, 38043 Grenoble, France}

% NOTES ON CODE
% CURRENT VERSION OF FILE:
% THIS CODE IS TAKEN FROM FILE nonspherical13.nb
% (WITH A FEW ADDITIONAL MODIFICATIONS)

% Code was taken using Mathematica's feature Save as TeX
% but I coloured input and output separately
% and then changed all SLASHpmbs to SLASHmathtts
% and all SLASHtexts to nothing
% then removed all DOLLAR signs from code
% and then changed all SLASHprimes to apostrophes

\maketitle

\begin{abstract}
Simple analytical formulae, directly relating the experimental geometry and sample orientation to the measured R(M)XS scattered intensity are very useful to design experiments and analyse data. Such formulae can be obtained by the contraction of an  expression containing the polarisations and crystal field tensors, and where the magnetisation vector acts as a rotation derivative\cite{mirone}. The result of a contraction contains a scalar product of (rotated) polarisation vectors and the crystal field axis. As an example, the dipole-dipole RXS scattering amplitude by Jahn-Teller distortion is calculated as:

\begin{eqnarray*}
\mathbb{C} ( \epsilon'(2z^2-x^2-y^2)\epsilon )
= 2
{\contraction{}{\epsilon'}{z}{z}
\contraction[0.5ex]{\epsilon'}{z}{z}{\epsilon}
\epsilon' z z \epsilon}
+ 2
{\contraction{}{\epsilon'}{z}{}
\epsilon' z}
{\contraction{}{z}{\epsilon}{}
z \epsilon}
+ ...
\\
= 4 \epsilon'_z \epsilon_z \newline - 2 \epsilon'_x \epsilon_x - 2 \epsilon'_y \epsilon_y
\end{eqnarray*}

The contraction rules give rise to combinatorial algorithms which can be efficiently treated by computers.
In this work we provide and discuss a concise Mathematica code along with a few example applications to non-centrosymmetric magnetic systems.
\end{abstract}

\section{Introduction}

A tensorial contraction method has been developed\cite{mirone} to obtain analytical formulae for X-ray resonant magnetic scattering, working from the established formulae for RMXS amplitude in the spherical atom approximation\cite{hm}. Here this method will be implemented in a Mathematica code and used to investigate electric dipole and quadrupole scattering in the general case of a non-spherical system.

To illustrate the capabilities of the code, we consider the case of \textit{spiral antiferromagnetic holmium}. This has an HCP structure and the atoms are embedded in a local $D_{3h}$ symmetry environment. The magnetic field direction of the holmium varies helically according to the crystal layer.

The Mathematica code executes contractions of the tensors of an incoming and outgoing photon with the field tensor of the crystal system, to obtain expressions for the dipole-dipole, quadrupole-quadrupole and mixed dipole-quadrupole corrections to the scattering amplitude for the spherical case.

\section{Theory}

\subsection{Spherical case}

We consider the interaction between matter and a photon described by a wave vector $\vec{k}$ and a polarisation vector $\vec{\epsilon}$, discarding both elastic Thompson scattering and the spin-magnetic field interaction. Our atom is spherical\cite{mirone} but perturbed by a magnetic exchange field. We take the angular momentum quantisation axis $\vec{\xi}$ along the magnetic field. Since dipolar and quadrupolar terms will not mix due to parity conservation, the scattering amplitude is given by\cite{mirone}:

\begin{equation}
\label{sphericaltensors}
F_{\epsilon,k \rightarrow \epsilon',k'} = \sum^{q=1}_{q=-1} F_{1,q}\epsilon^{1*}_q\epsilon^1_q + \sum^{q=2}_{q=-2} F_{2,q}(k\epsilon)^{*}_q(k\epsilon)_q
\end{equation}

where $(k\epsilon)_q$ denotes the rank 2 spherical tensor components of $\vec{k}\otimes\vec{\epsilon}$, and $\epsilon^1_q$ represents $\vec{\epsilon}$ in rank 1 spherical tensor form. The scattering coefficients $F_{1,q}$ and $F_{2,q}$ can be derived from the analysis in \cite{mirone}.

\subsection{Spherical case in Cartesian representation}

It is, however, more convenient to keep the vectors $\vec{k}$ and $\vec{\epsilon}$ in Cartesian form and re-express Equation~\ref{sphericaltensors} in Cartesian space. Using a different set of coefficients $F_1^{\prime n}$, $F_2^{\prime n}$ (which we will define in Section~\ref{definingfprefactors}), and substituting $q$ in Equation~\ref{sphericaltensors} by $\vec{\xi}.\vec{L}$ where $\vec{L}$ is the angular momentum operator, we re-express Equation~\ref{sphericaltensors} as
\begin{equation}
\label{fsphericalcartesian}
F_{\epsilon,k \rightarrow \epsilon',k'} = \mathbb{C}\left(\epsilon' \sum^{n=2}_{n=0} (i \xi \times)^n F^{\prime n}_1 \epsilon \right) + \mathbb{C}\left((k' \otimes \epsilon') \sum^{n=4}_{n=0} (i \xi \times)^n F^{\prime n}_2 (k \otimes \epsilon) \right) / 2
\end{equation}
where $\mathbb{C}$ signifies a sum over all possible contractions (the contractions are defined in Section~\ref{contractions}). This formulation can be used to calculate directly the dipolar and quadrupolar scattering coefficients for a spherical atom\cite{mirone}, which are in agreement with the formulae obtained previously by Hill \& McMorrow\cite{hm}. The implementation of $\mathbb{C}$ as a computer-efficient algorithm is one of the objectives of this code.

\subsection{Non-spherical case}

We use the extension of this formulation for the case of a non-spherical atom\cite{mirone}. We represent the non-sphericity of the atomic environment by a crystal potential $T(x,y,z)$ which is added to the atomic Hamiltonian, which is given as a superposition of spherical harmonics\cite{mirone}:
\begin{equation}
T(x,y,z) = \sum_{l,q} t_{l,q} T^l_q(x,y,z)
\end{equation}
where $T$ must also have the same symmetry as the crystal system\cite{ct}.

The field tensor must now be included in Equation~\ref{fsphericalcartesian}. The scattering amplitude for the non-spherical case is therefore given by the contraction sum

\begin{equation}
\label{fcartesiancontractions}
F_{\epsilon,k \rightarrow \epsilon',k'} = \mathbb{C}\left(\epsilon' \sum^{n=2}_{n=0} (i \xi \times)^n F^{\prime n}_1 T \epsilon \right) + \mathbb{C}\left((k' \otimes \epsilon') \sum^{n=4}_{n=0} (i \xi \times)^n F^{\prime n}_2 T (k \otimes \epsilon) \right) / 2
\end{equation}

Later we will construct some code to evaluate $F_{\epsilon,k \rightarrow \epsilon',k'}$ in the general case of a field tensor in $x$, $y$ and $z$, and then investigate the effect that this will have on the scattering peaks measured.

\subsection{Contractions}
\label{contractions}

First we define mathematically what is meant by a \textit{contraction}\cite{snieder} and the function $\mathbb{C}$, which represents the sum of all possible contractions between the tensors it is operating on, that result in a scalar. In Section~\ref{program} this will be implemented as a computer code.

\paragraph{Single contractions}

A single contraction of two tensors involves taking the inner product between an index of the first tensor $\vec{a}_1 \otimes \vec{a}_2$ and one of the second tensor $\vec{b}_1 \otimes \vec{b}_2$, thus reducing the total rank of the expression by two. So, for two tensors $\vec{a}_1 \otimes \vec{a}_2$ and $\vec{b}_1 \otimes \vec{b}_2$, each of rank 2, the possible single contractions are
\begin{equation}
\label{singlecontraction}
(\vec{a}_1 . \vec{b}_1) \vec{a}_2 \vec{b}_2
\;\;\;\;
(\vec{a}_1 . \vec{b}_2) \vec{a}_2 \vec{b}_1
\;\;\;\;
(\vec{a}_2 . \vec{b}_1) \vec{a}_1 \vec{b}_2
\;\;\;\;
(\vec{a}_2 . \vec{b}_2) \vec{a}_1 \vec{b}_1
\end{equation}
i.e. there are four possible single contractions of $\vec{a}_1 \otimes \vec{a}_2$ with $\vec{b}_1 \otimes \vec{b}_2$, each one a tensor of rank 2.

\paragraph{Double contractions}

However, we are interested in contracting the two tensors to form a scalar. For the case of the two rank-2 tensors $\vec{a}_1 \otimes \vec{a}_2$ and $\vec{b}_1 \otimes \vec{b}_2$, we can execute the first single contraction and then contract the two indices that have not already been contracted. Working from Equation~\ref{singlecontraction}, we see that this results in a total of four possible (double) contractions:
\begin{equation}
\label{doublecontraction}
(\vec{a}_1 . \vec{b}_1) (\vec{a}_2 . \vec{b}_2)
\;\;\;\;
(\vec{a}_1 . \vec{b}_2) (\vec{a}_2 . \vec{b}_1)
\;\;\;\;
(\vec{a}_2 . \vec{b}_1) (\vec{a}_1 . \vec{b}_2)
\;\;\;\;
(\vec{a}_2 . \vec{b}_2) (\vec{a}_1 . \vec{b}_1)
\end{equation}

Therefore, the contraction sum of $\vec{a}_1 \otimes \vec{a}_2$ and $\vec{b}_1 \otimes \vec{b}_2$, which is the sum of all possible contractions, is
\begin{equation}
\mathbb{C}(A_{ij}B_{kl}) = 2 (\vec{a}_1 . \vec{b}_1) (\vec{a}_2 . \vec{b}_2) + 2 (\vec{a}_2 . \vec{b}_1) (\vec{a}_1 . \vec{b}_2)
\end{equation}
Note the factors of 2 which arise from the repeated terms in Equation~\ref{doublecontraction}.

\paragraph{Multiple contractions}

We would like to construct an iterative computer algorithm to take a set of tensors and repeatedly contract them until the result becomes a scalar.

We begin by taking all possible single contractions (for the case of two tensors of rank $N$, there are $2N$ possible single contractions; however we would also like to contract three tensors with each other).

Then we take this result and take all possible single contractions on it again---this continues until either the expression becomes a scalar (in which case the contraction function is no longer called, and the result is output as the answer), or a tensor of rank 1 (which will be ignored by the program, as it cannot be contracted to a scalar).

\section{The program}
\label{program}

Here we describe the workings of the Mathematica notebook. The program input is presented in a different colour for easy differentiation from the rest of the document. Occasionally the program output has been stylised slightly for this report, but everything listed here was produced more or less directly by Mathematica.

In this section we will define the contraction mechanisms for calculating the magnetic diffraction amplitude for the general case of a material with a known field tensor.

In Section~\ref{spiralantiferromagneticholmium} we will then apply this mechanism for the specific case of spiral antiferromagnetic holmium, and show how this can be used to calculate the expected intensities of the satellite diffraction peaks of Bragg order $n \pm q$.

\subsection{Tensor contraction}

First we set up a function to execute tensorial contractions (these are needed to calculate the scattering amplitudes for a non-spherical atom).

\paragraph{Contracting two tensors}

First we define a function $\mathtt{Con}$ which will find the sum of all possible contractions for two tensors rank $N$. Each tensor is given in the form $\mathtt{T}[\vec{a}_1, \vec{a}_2 ... \vec{a}_N]$ (representing the tensor product $\vec{a}_1 \otimes \vec{a}_2 \otimes ... \otimes \vec{a}_N$ of a set of vectors $\vec{a}_1$, $\vec{b}_2$, ... $\vec{a}_N$).
For example, the contraction sum of two tensors $\vec{a}_1 \otimes \vec{a}_2$ and $\vec{b}_1 \otimes \vec{b}_2$ will be entered as $\mathtt{Con}[\mathtt{T}[\vec{a}_1,\vec{a}_2], \mathtt{T}[\vec{b}_1,\vec{b}_2]]$, and is equivalent to, mathematically,
\begin{equation}
\mathbb{C}((\vec{a}_1 \otimes \vec{a}_2) (\vec{b}_1 \otimes \vec{b}_2)) = (\vec{a_1}.\vec{b}_1)(\vec{a}_2.\vec{b}_2) + (\vec{a}_1.\vec{b}_2)(\vec{a}_2.\vec{b}_1)
\end{equation}
i.e. each vector in the left hand tensor is contracted with each one on the right, thus generating all possible contractions.

The contraction function $\mathtt{Con}$ for two tensors is given below. It loops round, gradually reducing the expression by eliminating pairs of vectors and calling itself again to contract the remaining expression. This creates a complicated stack of one function calling itself again many times with different arguments---to prevent infinite loops in the case of an error, in this definition $\mathtt{Con}$ will not call itself directly but will call a temporary (as yet undefined) function $\mathtt{ConTmp}$ which can later be set equal to $\mathtt{Con}$.

First we state that the function $\mathtt{Con}$ and our tensor product operator $\mathtt{T}$ are both \textit{orderless}, so it does not matter in which order their arguments (vectors and tensors) are given.\\

\noindent\(\color{code}\mathtt{{Attributes}[{Con}]=\{{Orderless}\};}\\
\mathtt{{Attributes}[T] = \{{Orderless}\};}\\\)

Now we create the contraction algorithm. The function takes two tensors and returns the sum of all possible single contractions between them.\\

\noindent\(\color{code}\mathtt{{Con}[{x\_\_},0]{:=}0;}\\
\mathtt{{Con}[T[{a\_}, {al\_\_\_}],T[{bl\_\_\_}]]{:=}{Module}[\{{res}\},{res}=0;}\\
\mathtt{~~n={Length}[{List}[{bl}]];}\\
\mathtt{~~{Do}[}\\
\mathtt{~~{res}=}\\
\mathtt{~~~~{res} + {scal}\left[a{/.}\left\{x\to e_1,y\to e_2,z\to e_3\right\},\right.}\\
\mathtt{~~~~\left.{List}[{bl}][[i]]{/.}\left\{x\to e_1,y\to e_2,z\to e_3\right\}\right] }\\
\mathtt{~~~~\times {ConTmp}[ T[{al}], {Delete}[ T[{bl}],i] ]}\\
\mathtt{~~,\{i,1,n\}}\\
\mathtt{~~];}\\
\mathtt{~~{res}}\\
\mathtt{];}\)
\\

\noindent where we introduce the variables $\vec{e}_1$, $\vec{e}_2$ and $\vec{e}_3$ to represent the basis vectors in the $\vec{\hat{x}}$, $\vec{\hat{y}}$ and $\vec{\hat{z}}$ directions (this will be useful later for when we will need to take scalar products).

Some other definitions are necessary to enable the program to stop looping when it has finished the contractions. The contraction of a single tensor must evaluate to zero; to contract anything with 1, we can ignore the 1; and the contraction of nothing must evaluate to 1.
\\

\noindent\(\color{code}\mathtt{{Con}[T[{c\_\_\_}]]{:=}0;}\\
\mathtt{{Con}[1,{t\_\_\_}]{:=}{Con}[t];}\\
\mathtt{{Con}[]{:=}1;}\\
\mathtt{{  }T[]=1; }\\
\mathtt{T[\{\}]=1;}\\
\mathtt{T[\{{x\_\_}\}]=T[x];}\)

\paragraph{Contracting three tensors}

Now a contraction function has been defined for two tensors, we can start to think about $\mathbb{C}$ for three tensors. This is more complicated, as the sum of possible contractions is much greater. The function works in the same way as the $\mathtt{Con}$ function for the case of two tensors, eliminating vector pairs one by one and calling itself (via $\mathtt{ConTmp}$) repeatedly. Once one of the three tensors has been fully contracted with vector elements from the other two, the expression will have been reduced to a sum of contractions of \textit{two} tensors---which is a much simpler problem and has already been defined above.\\

\noindent\(\color{code}\mathtt{{Con}[T[{al\_\_\_}],T[{bl\_\_\_}],T[{cl\_\_\_}]]{:=}}\\
\mathtt{{Module}[\{{res},m,n\},{res}=0;}\\
\mathtt{~~m={Length}[T[{bl}]];}\\
\mathtt{~~n={Length}[T[{cl}]];}\\
\mathtt{~~{Do}[}\\
\mathtt{~~~~{res}=}\\
\mathtt{~~~~~~{res} + {scal}\left[{List}[{al}][[1]]{/.}\left\{x\to e_1,y\to e_2,z\to e_3\right\},\right.}\\
\mathtt{~~~~~~\left.{List}[{bl}][[i]]{/.}\left\{x\to e_1,y\to e_2,z\to e_3\right\}\right] }\\
\mathtt{~~~~~~\times {ConTmp}[ {Delete}[ T[{al}],1], {Delete}[ T[{bl}],i],T[{cl}] ]}\\
\mathtt{~~~~,\{i,1,m\}}\\
\mathtt{~~~~];}\)
\noindent\(\color{code}\mathtt{~~~~{Do}[}\\
\mathtt{~~~~{res}=}\\
\mathtt{~~~~~~{res} + {scal}\left[{List}[{al}][[1]]{/.}\left\{x\to e_1,y\to e_2,z\to e_3\right\},\right.}\\
\mathtt{~~~~~~\left.{List}[{cl}][[i]]{/.}\left\{x\to e_1,y\to e_2,z\to e_3\right\}\right] }\\
\mathtt{~~~~~~\times {ConTmp}[ {Delete}[ T[{al}],1],T[{bl}],{Delete}[ T[{cl}],i] ]}\\
\mathtt{~~~~,\{i,1,n\}}\\
\mathtt{~~~~];}\\
\mathtt{~~~~{res}}\\
\mathtt{~~];}\)
\\

\noindent As a temporary measure, we also make all callings of $\mathtt{ConTmp}$ equivalent to calling the function $\mathtt{Con}$ itself (thus creating a stack)---in the case of errors or infinite loops, this line can be removed.\\

\noindent\(\color{code}\mathtt{{ConTmp}={Con}}\)

\subsection{Extending the contraction functions to work with polynomials}

\paragraph{The field tensor in polynomial representation}

Now we need to make a function which will contract a Cartesian tensor in the form of a polynomial in $x$, $y$ and $z$. This is needed for scattering amplitude calculations, where the crystal field tensor is typically expressed in a polynomial form $T = f(x, y, z)$. For example, the quadrupole-quadrupole scattering formula for the particular case of holmium has the form:

\begin{equation}
F_{\epsilon, k \rightarrow \epsilon', k'}^{q,q} = \mathbb{C}((\epsilon' \otimes k') (t_2(2z^2-x^2-y^2) \pm t_3(x^3-3x y^2)) (\epsilon \otimes k))
\end{equation}

We define a new function $\mathbb{C}(A, f(x, y, z), C)$ which re-expresses a contraction of tensors $A$ and $C$ and the polynomial $f$ in terms of sums and products of contractions of three tensors which can be executed by the function $\mathtt{Con}$.

\paragraph{Reducing the polynomial via linearity}

We want to be able to contract some quite complicated polynomials, so it is useful to first reduce the polynomials to a more manageable form using some basic mathematical properties of the function $\mathbb{C}$. For example, the contraction sum operation is linear, so

\begin{equation}
\mathbb{C}(A (f(x, y, z) + g(x, y, z)) B) = \mathbb{C}(A f(x, y, z) B) + \mathbb{C}(A g(x, y, z) B)
\end{equation}

So we can start constructing the function $\mathbb{C}$ by creating a \texttt{Rule}\footnote{In Mathematica, a \texttt{Rule} defining $\mathbb{C}$ means that every time $\mathbb{C}$ is referred to in subsequent code, the code within the \texttt{Rule} is invoked. Parameters on the left side of a definition of a \texttt{Rule} must be followed by underscores (e.g. \texttt{A\_}), to show that they can take any value.} that any polynomial to be contracted must first be split into terms and the final contraction will be a sum of contractions of these terms:\\

\noindent\(\color{code}\mathtt{\mathbb{C}[\{{al\_\_}\},{Plus}[{x\_},{y\_\_}],\{{cl\_\_}\}]{:=}}\\
\mathtt{~~{Module}[\{i\},{xy}={Expand}[x+y];}\\
\mathtt{~~~~{Sum}[\mathbb{C}[\{{al}\},{xy}[[i]],\{{cl}\}],\{i,1,{Length}[{xy}]\}]]}\)

\paragraph{Removing prefactors from the individual terms}

Now we need to find an efficient way of taking each term and extracting the $xyz$ dependent components for contraction, while leaving numbers and constants untouched by the $\mathtt{Con}$ function. Here we use another property of the linearity of $\mathbb{C}$:

\begin{equation}
\mathbb{C}(A (k f(x, y, z)) B) = k \mathbb{C}(A f(x, y, z) B) 
\end{equation}

So first we separate out any occurrences of $xyz$ from the terms of the polynomial (which has already been split up into a sum of parts):\\

\noindent\(\color{code}\mathtt{\mathbb{C}[\{{al\_\_}\},}\\
\mathtt{~~{Times}[{prefactor\_}{/;}{FreeQ}[{FullForm}[{prefactor}],x]{/;}}\\
\mathtt{~~{FreeQ}[{FullForm}[{prefactor}],y]{/;}}\\
\mathtt{~~{FreeQ}[{FullForm}[{prefactor}],z],{xyz\_\_}],\{{cl\_\_}\}]{:=}}\\
\mathtt{~~{Times}[{prefactor}] \mathbb{C}[\{{al}\},{Times}[{xyz}],\{{cl}\}];}\)

\paragraph{Converting a polynomial term to a tensor}

Now that individual terms in $xyz$ have been separated, they can be converted into a tensorial form (such as $\mathtt{T[x,x,y,z]}$ in Mathematica) that can be used directly by the function $\mathtt{Con}$.

\paragraph{Expressing $x^\alpha y^\beta z^\gamma$ in tensorial form}

First we make a \texttt{Rule} that $\mathbb{C}(A (x^\alpha y^\beta z^\gamma) B)$ will be executed as $\mathtt{Con[T[A],T[B],T[C]]}$ where $\mathtt{B}$ is a sequence of the variables $\mathtt{x}$, $\mathtt{y}$ and $\mathtt{z}$. For example, we want the expression
\begin{equation}
\mathtt{\mathbb{C}[T[A], (x^2 y z), T[C]]} \equiv \mathbb{C}(A (x^2 y z) B) \mathit{~(mathematical~notation)}
\end{equation}
to be converted to
\begin{equation}
\mathtt{Con[T[A],T[x,x,y,z],T[C]]}
\end{equation}
and then contracted in the usual way by the function $\mathtt{Con}$. The code to execute this is as follows:\\

\noindent\(\color{code}\mathtt{\mathbb{C}[\{{al\_\_}\},{Times}[{xyz1\_},{xyz2\_\_}],\{{cl\_\_}\}]{:=} }\\
\mathtt{~~{Module}[\{i,{power},n,{xyz}\},}\\
\mathtt{~~~~{xyz}{:=}{List}[{xyz1},{xyz2}]{/.}{x\_}^{{power\_}}{->}{Table}[x,\{i,1,{power}\}];}\\
\mathtt{~~~~n={Length}[{xyz}];}\\
\mathtt{~~~~{Con}[T[{al}],T[{Flatten}[{Table}[{xyz}[[i]],\{i,1,n\}]]],T[{cl}]]}\\
\mathtt{~~]}\)

\paragraph{Expressing $x^\alpha$ in tensorial form}
We also need a \texttt{Rule} to interpret powers of a single variable, such as $\mathbb{C}(A (x^\alpha) B)$, since this case is not covered by the above \texttt{Rule} (although the interpretation is much the same).\\

\noindent\(\color{code}\mathtt{\mathbb{C}[\{{al\_\_}\},{Power}[{xyz\_},{power\_}],\{{cl\_\_}\}]{:=} }\\
\mathtt{~~{Module}[\{i,n\},}\\
\mathtt{~~~~{Con}[T[{al}],T[{Table}[{xyz},\{i,1,{power}\}]],T[{cl}]]}\\
\mathtt{~~]}\)

\subsection{Defining the scalar product symbolically}

The contraction function defined so far has produced results in terms of the function $\mathtt{scal}[\vec{a},\vec{b}]$, as yet undefined. We must now define $\mathtt{scal}$ so that a product of a Cartesian vector with one of the basis vectors $\hat{\mathbf{x}}$, $\hat{\mathbf{y}}$ and $\hat{\mathbf{z}}$ will be re-expressed as the correct component of that vector (at least for the moment).\\

\noindent\(\color{code}\mathtt{{Clear}[{scal},\xi ];}\\
\mathtt{{Attributes}[{scal}]=\{{Orderless}\};}\\
\mathtt{{scal}\left[{vector\_},e_{{i\_}}\right]{:=}{vector}.\{\hat{\mathbf{x}},\hat{\mathbf{y}},\hat{\mathbf{z}}\}[[i]];}\)

\section{Application: spiral antiferromagnetic holmium}
\label{spiralantiferromagneticholmium}

The contraction mechanism which has been defined will work in general on any kind of polynomial tensor. Now we present an example of its application, for the specific case of spiral antiferromagnetic holmium

\subsection{Defining the geometry of the system}

\paragraph{Defining frame $X$}

\begin{figure}
\label{planegeometry}
\includegraphics[width=8cm]{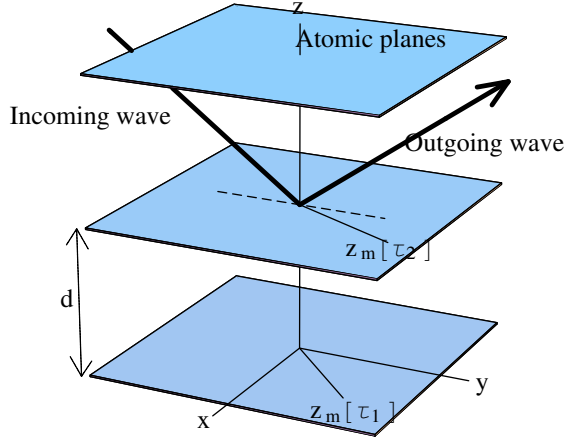}
\caption{The geometry of the holmium system. Since $\tau$ is different for different crystal planes, $\vec{\hat{z}}_m$ has a spiral dependence on $z$.}
\end{figure}

First, we define a simple Cartesian co-ordinate system. Initially we would like to work in the frame $X$ with co-ordinates $(x, y, z)$ oriented around the crystal geometry ($z$ points upwards through the crystal planes), as shown in Figure~\ref{planegeometry}.

The magnetic field direction $\vec{\hat{z}}_m$ of spiral antiferromagnetic holmium varies helically  according to the crystal layering, as 
\begin{equation}
\vec{\hat{z}}_m(\tau) = (\cos \tau, \sin \tau, 0)
\end{equation}
where the angle $\tau = \frac{2 \pi q z}{d}$ depends on the height $z$ ($d$ is the interplanar spacing and $q$ is the antiferromagnetic wavevector). In Mathematica code this is expressed as (see Figure~\ref{planegeometry}):\\

\noindent\(\color{code}\mathtt{\hat{z}_m[\tau \_]=\{{Cos}[\tau ],{Sin}[\tau ],0\};}\)

\paragraph{The incoming/outcoming photons}

Now we define the characteristics in frame $X$ of an incoming photon, at angle $\theta$ from the crystal planes. Its wavevector is\\

\noindent\(\color{code}\mathtt{{kx}[\theta \_]=\{{Cos}[\theta ],0,-{Sin}[\theta ]\};}\)\\

and its polarisation vectors in $X$ are (with the indices 0 and 1 representing polarisation orientations $\sigma$ and $\pi$ respectively):\\

\noindent\(\color{code}\mathtt{{\epsilon x}_0[\theta \_]=\{0,1,0\};}\)

\noindent\(\color{code}\mathtt{{\epsilon x}_1[\theta \_]=\{{Sin}[\theta ],0,{Cos}[\theta ]\};}\)

\noindent\(\color{code}\mathtt{{vector\_}_{\sigma }{:=}{vector}_0;}\)

\noindent\(\color{code}\mathtt{{vector\_}_{\pi }{:=}{vector}_1;}\)

\subsection{The magnetic co-ordinate system}

\paragraph{Defining $M$}

It is more convenient for our purposes to introduce a new co-ordinate system $M$ with basis vectors $(\vec{\hat{x}}_m, \vec{\hat{y}}_m, \vec{\hat{z}}_m)$ here, constructed from the (rotating) magnetic moment $\vec{\hat{z}}_m$. We set $\vec{\hat{x}}_m(\tau)$ to the most convenient possibility:\\

\noindent\(\color{code}\mathtt{\hat{x}_m[\tau \_]=\{0,0,1\};}\)\\

\noindent And we can now derive $\vec{\hat{y}}_m(\tau)$ from $\vec{\hat{x}}_m(\tau)$ and $\vec{\hat{z}}_m(\tau)$:\\

\noindent\(\color{code}\mathtt{\hat{y}_m[\tau \_]=\hat{x}_m[\tau ]\times \hat{z}_m[\tau ];}\)\\

\noindent So our new basis is

\begin{equation}
\left( \vec{\hat{x}}_m | \vec{\hat{y}}_m | \vec{\hat{z}}_m \right)
=
\left(
\begin{array}{lll}
 0 & -\sin \tau & \cos \tau \\
 0 & \cos \tau & \sin \tau \\
 1 & 0 & 0
\end{array}
\right)
\end{equation}

\paragraph{Re-expressing wave parameters in the magnetic co-ordinate system}

Now we can redefine $\vec{k}$ and $\vec{\epsilon}_{\pi,\sigma}$ in the frame $M$ from their definitions $\mathtt{kx}$ and $\mathtt{{\epsilon}x}_{\pi,\sigma}$ in frame $X$ (the subscript $\mathtt{pol}$ stands for either $\pi$ or $\sigma$):\\

\noindent\(\color{code}\mathtt{k[\theta \_,\tau \_]{:=}\left\{{kx}[\theta ].\hat{x}_m[\tau ],{kx}[\theta ].\hat{y}_m[\tau ],{kx}[\theta ].\hat{z}_m[\tau
]\right\};}\)

\noindent\(\color{code}\mathtt{\epsilon _{{pol\_}}[\theta \_,\tau \_]{:=}}
\mathtt{\left\{{\epsilon x}_{{pol}}[\theta ].\hat{x}_m[\tau ],{\epsilon x}_{{pol}}[\theta ].\hat{y}_m[\tau
],\right.}
\mathtt{\left.{\epsilon x}_{{pol}}[\theta ].\hat{z}_m[\tau ]\right\};}\)\\

\noindent So the wave vector in this new basis is
\begin{equation}
\label{kdefinition}
\color{output}
\vec{k}(\theta ,\tau) =
\left(
\begin{array}{l}
 -\sin (\theta ) \\
 -\cos (\theta ) \sin (\tau ) \\
 \cos (\theta ) \cos (\tau )
\end{array}
\right)
\end{equation}
And the polarisation vectors are
\begin{equation}
\label{epsilondefinition}
\color{output}
\vec{\epsilon}_{\sigma }(\theta ,\tau) =
\left(
\begin{array}{l}
 0 \\
 \cos (\tau ) \\
 \sin (\tau )
\end{array}
\right)
;
\vec{\epsilon}_{\pi }(\theta ,\tau) =
\left(
\begin{array}{l}
 \cos (\theta ) \\
 -\sin (\theta ) \sin (\tau ) \\
 \cos (\tau ) \sin (\theta )
\end{array}
\right)
\end{equation}

\paragraph{The reflected wave}

We define the reflected wave parameters $\vec{k}'$, $\vec{\epsilon}'$ by changing the sign of $\theta$:\\

\noindent\(\color{code}\mathtt{\epsilon' _{{pol\_}}[\theta \_,\tau \_]{:=}\epsilon _{{pol}}[-\theta ,\tau ];}\)

\noindent\(\color{code}\mathtt{{k' }[\theta \_,\tau \_]{:=}k[-\theta ,\tau ];}\)

\subsection{Applying the contraction mechanism for the Holmium case}

Using our expression for the Holmium field tensor (ignoring the $\pm$ alternation),
\begin{equation}
T = t2(2z^2-x^2-y^2)+t_3(x^3-3 x y^2)
\end{equation}
we can use the functions $\mathtt{\mathbb{C}}$ and $\mathtt{Con}$ to derive some properties of the diffraction pattern, that arise from the non-sphericity of the holmium atom. We are interested in particular in effects that are unique for the non-spherical case---\textit{corrections} to the spherical scattering formulae are what we are after. We will analyse separately the dipole-dipole, quadrupole-quadrupole and mixed terms, as the whole expression is rather complicated.

\paragraph{Calculating the dipole-dipole scattering correction}

First we look at the simplest case: the dipole-dipole scattering correction. Here we only consider the contraction of $\epsilon'$ and $\epsilon$ (the outgoing and incoming photon polarisation tensors respectively) with $T$, ignoring $k'$ and $k$: that is, we ignore effects of the finiteness of the speed of light. We will attempt to find this function's proportionality (i.e. we are not interested in prefactors for now).

The contraction is (taking only the first order term in the operator $(i \xi \times)^n$):

\begin{equation}
\mathbb{C}(\epsilon' T (i \xi \times \epsilon)) - [\epsilon' \leftrightarrow \epsilon]
\end{equation}

To encode this contraction in Mathematica, we ignore for now the symmetry term $-[\epsilon' \leftrightarrow \epsilon]$, and enter:\\

\noindent\(\color{code}\mathtt{\mathbb{C}\left[\{\epsilon' \},{t2}\left(2z^2-x^2-y^2\right)+{t3}\left(x^3-3 x y^2\right),\{i \xi \times \epsilon
\}\right]{/.}{t2}\to 1}\)

\begin{equation}
\color{output}
-2 \epsilon' .\hat{\mathbf{x}} (i \xi \times \epsilon ).\hat{\mathbf{x}}-2 \epsilon' .\hat{\mathbf{y}} (i \xi \times \epsilon ).\hat{\mathbf{y}}+4
\epsilon' .\hat{\mathbf{z}} (i \xi \times \epsilon ).\hat{\mathbf{z}}
\end{equation}

Condensing the spherical components into a single term gives (still ignoring the symmetry term)\\

\noindent\(\color{code}\mathtt{{Simplify}[\%,(\epsilon' .\hat{\mathbf{x}} (i \xi \times \epsilon ).\hat{\mathbf{x}}+\epsilon' .\hat{\mathbf{y}} (i \xi \times \epsilon
).\hat{\mathbf{y}}+ \epsilon' .\hat{\mathbf{z}} (i \xi \times \epsilon ).\hat{\mathbf{z}})==}\\
\mathtt{\epsilon' .(i \xi \times \epsilon )]}\)

\begin{equation}
\color{output}
-2 \epsilon' .(i \xi \times \epsilon )+6 \epsilon' .\hat{\mathbf{z}} (i \xi \times \epsilon ).\hat{\mathbf{z}}
\end{equation}

Subtracting the symmetry term $[\epsilon' \leftrightarrow \epsilon]$, and removing constant prefactors, we see that the dipole-dipole scattering correction is proportional to:\\

\noindent\(\color{code}\mathtt{{Simplify}[\%/2-{symm}]{//}{TraditionalForm}}\)

\begin{equation}
\color{output}
-{symm}-\epsilon' .(i \xi \times \epsilon )+3 \epsilon' .\hat{\mathbf{z}} (i \xi \times \epsilon ).\hat{\mathbf{z}}
\end{equation}

This is Equation~29 in \cite{mirone}. The $\epsilon' .(i \xi \times \epsilon )$ part merges with the term that appears in the spherical formulae, but the other term adds complexity to the amplitude dependence on experimental geometry.

\paragraph{Calculating the quadrupole-quadrupole scattering correction}

Mathematically, the contraction we are interested in is

\begin{equation}
\begin{array}{c}
\mathbb{C}((k' \otimes \epsilon') T (i \xi \times) (k \otimes \epsilon)) - [k',\epsilon' \leftrightarrow k,\epsilon] \\
= \left[\mathbb{C}((k' \otimes \epsilon') T ((i \xi \times k) \otimes \epsilon))
+ \mathbb{C}((k' \otimes \epsilon') T (k \otimes (i \xi \times \epsilon)))\right] \\
 - [k',\epsilon' \leftrightarrow k,\epsilon]
\end{array}
\end{equation}

In Mathematica, this is encoded as (ignoring the $\{i \xi \times k,\epsilon \}$ term, and also the symmetry terms, for now):\\

\noindent\(\color{code}\mathtt{\mathbb{C}\left[\{{k' },\epsilon' \},{t2}\left(2z^2-x^2-y^2\right)+{t3}\left(x^3-3 x y^2\right),\{k,i
\xi \times \epsilon \}\right]{/.}}\\
\mathtt{{t2}\to 1}\)\\

\noindent which gives output

\begin{equation}
\begin{array}{c}
\color{output}-2 \epsilon' .\hat{\mathbf{x}} (i \xi \times \epsilon ).\hat{\mathbf{x}} \mathtt{~scal}[k,{k' }]-2 \epsilon' .\hat{\mathbf{y}} (i \xi
\times \epsilon ).\hat{\mathbf{y}} \mathtt{~scal}[k,{k' }]-\\
\color{output}2 {k' }.\hat{\mathbf{x}} (i \xi \times \epsilon ).\hat{\mathbf{x}} \mathtt{~scal}[k,\epsilon' ]-2 {k' }.\hat{\mathbf{y}} (i \xi \times \epsilon
).\hat{\mathbf{y}} \mathtt{~scal}[k,\epsilon' ]-\\
\color{output}2 k.\hat{\mathbf{x}} (\epsilon' .\hat{\mathbf{x}} \mathtt{~scal}[{k' },i \xi \times \epsilon ]+{k' }.\hat{\mathbf{x}} \mathtt{~scal}[\epsilon
' ,i \xi \times \epsilon ])-\\
\color{output}2 k.\hat{\mathbf{y}} (\epsilon' .\hat{\mathbf{y}} \mathtt{~scal}[{k' },i \xi \times \epsilon ]+{k' }.\hat{\mathbf{y}} \mathtt{~scal}[\epsilon
' ,i \xi \times \epsilon ])+\\
\color{output}2 (2 \epsilon' .\hat{\mathbf{z}} (i \xi \times \epsilon ).\hat{\mathbf{z}} \mathtt{~scal}[k,{k' }]+2 {k' }.\hat{\mathbf{z}} (i \xi \times
\epsilon ).\hat{\mathbf{z}} \mathtt{~scal}[k,\epsilon' ]+\\
\color{output}2 k.\hat{\mathbf{z}} (\epsilon' .\hat{\mathbf{z}} \mathtt{~scal}[{k' },i \xi \times \epsilon ]+{k' }.\hat{\mathbf{z}} \mathtt{~scal}[\epsilon
' ,i \xi \times \epsilon ]))
\end{array}
\end{equation}

Condensing the spherical components into a single term,
\\

\noindent\(\color{code}\mathtt{{Simplify}[\%,}\\
\mathtt{\{\epsilon' .\hat{\mathbf{x}} (i \xi \times \epsilon ).\hat{\mathbf{x}}+\epsilon' .\hat{\mathbf{y}} (i \xi \times \epsilon ).\hat{\mathbf{y}}+ \epsilon
' .\hat{\mathbf{z}} (i \xi \times \epsilon ).\hat{\mathbf{z}}==\epsilon' .(i \xi \times \epsilon ),}\\
\mathtt{{k' }.\hat{\mathbf{x}} (i \xi \times \epsilon ).\hat{\mathbf{x}}+{k' }.\hat{\mathbf{y}} (i \xi \times \epsilon ).\hat{\mathbf{y}}+ {k'
}.\hat{\mathbf{z}} (i \xi \times \epsilon ).\hat{\mathbf{z}}=={k' }.(i \xi \times \epsilon ),}\\
\mathtt{k.\hat{\mathbf{x}} \epsilon' .\hat{\mathbf{x}}+k.\hat{\mathbf{y}} \epsilon' .\hat{\mathbf{y}}+ k.\hat{\mathbf{z}} \epsilon' .\hat{\mathbf{z}}==k.\epsilon'
,}\\
\mathtt{k.\hat{\mathbf{x}} {k' }.\hat{\mathbf{x}}+k.\hat{\mathbf{y}} {k' }.\hat{\mathbf{y}}+ k.\hat{\mathbf{z}} {k' }.\hat{\mathbf{z}}==k.{k'
}\}]}\)

\begin{equation}
\begin{array}{c}
\color{output}-2 (\epsilon' .(i \xi \times \epsilon ) \mathtt{~scal}[k,{k' }]+{k' }.(i \xi \times \epsilon ) \mathtt{~scal}[k,\epsilon
' ]-\\
\color{output}3 {k' }.\hat{\mathbf{z}} (i \xi \times \epsilon ).\hat{\mathbf{z}} \mathtt{~scal}[k,\epsilon' ]+k.\epsilon'  \mathtt{~scal}[{k'
},i \xi \times \epsilon ]-\\
\color{output}3 \epsilon' .\hat{\mathbf{z}} ((i \xi \times \epsilon ).\hat{\mathbf{z}} \mathtt{~scal}[k,{k' }]+k.\hat{\mathbf{z}} \mathtt{~scal}[{k' },i
\xi \times \epsilon ])+\\
\color{output}k.{k' } \mathtt{~scal}[\epsilon' ,i \xi \times \epsilon ]-3 k.\hat{\mathbf{z}} {k' }.\hat{\mathbf{z}} \mathtt{~scal}[\epsilon'
,i \xi \times \epsilon ])
\end{array}
\end{equation}

\noindent and now taking only the non-spherical terms in $F$,
\\

\noindent\(\color{code}\mathtt{{Simplify}[\%{/.}\{\epsilon' .(i \xi \times \epsilon )\to 0,{k' }.(i \xi \times \epsilon )\to 0,k.\epsilon
' \to 0,k.{k' }\to 0\}]}\)

\begin{equation}
\begin{array}{c}
\color{output}6 (\epsilon' .\hat{\mathbf{z}} ((i \xi \times \epsilon ).\hat{\mathbf{z}} \mathtt{~scal}[k,{k' }]+k.\hat{\mathbf{z}} \mathtt{~scal}[{k'
},i \xi \times \epsilon ])+\\
\color{output}{k' }.\hat{\mathbf{z}} ((i \xi \times \epsilon ).\hat{\mathbf{z}} \mathtt{~scal}[k,\epsilon' ]+k.\hat{\mathbf{z}} \mathtt{~scal}[\epsilon' ,i \xi
\times \epsilon ]))
\end{array}
\end{equation}

Adding the $\{i \xi \times k,\epsilon \}$ term back in now (by exchanging $k$ and $i \xi \times \epsilon$ and adding the result on to the original formula), and rewriting $\mathtt{scal}$ in the dot-product notation, we find that the quadrupole-quadrupole scattering correction is proportional to:\\

\noindent\(\color{code}\mathtt{{Simplify}[\%+(\%{/.}k\to {tmp}{/.}\epsilon \to k{/.}{tmp}\to \epsilon )]/6{/.}}\\
\mathtt{{scal}[{vec1\_},{vec2\_}]\to {vec1}.{vec2}{//}{TraditionalForm}}\)

\begin{equation}
\begin{array}{c}
\color{output}\epsilon' .\hat{\mathbf{z}} (k.\hat{\mathbf{z}} {k' }.(i \xi \times \epsilon )+\\
\color{output}{k' }.(i \xi \times k) \epsilon .\hat{\mathbf{z}}+{k' }.\epsilon  (i \xi \times k).\hat{\mathbf{z}}+k.{k' } (i \xi \times
\epsilon ).\hat{\mathbf{z}})+\\
\color{output}{k' }.\hat{\mathbf{z}} (\epsilon .\hat{\mathbf{z}} \epsilon' .(i \xi \times k)+k.\hat{\mathbf{z}} \epsilon' .(i \xi \times \epsilon )+\\
\color{output}\epsilon .\epsilon'  (i \xi \times k).\hat{\mathbf{z}}+k.\epsilon'  (i \xi \times \epsilon ).\hat{\mathbf{z}})
\end{array}
\end{equation}

\noindent which is equivalent to Equation~30 in \cite{mirone}.

\paragraph{Calculating the mixed (quadrupole-dipole) scattering correction}

For the quadrupole-quadrupole and dipole-dipole scattering corrections calculated above, only the $t_2$ (quadratic order) terms of $T$ had any effect. The alternating term $\pm t_3(x^3-3 x y^2)$ contributes to the quadrupole-dipole and dipole-quadrupole scattering factor corrections, due to its cubic order.

The mixed scattering correction is (ignoring for now the extra terms in $[\epsilon \leftrightarrow k]$ and $[k \leftrightarrow k', \epsilon \leftrightarrow \epsilon']$):\\

\noindent\(\color{code}\mathtt{{Expand}\left[\mathbb{C}\left[\{\epsilon' \},{t2}\left(2z^2-x^2-y^2\right)+{t3}\left(x^3-3 x y^2\right),\{i
\xi \times k,\epsilon \}\right]{/.}\right.}\\
\mathtt{\{{t2}\to 0,{t3}\to 1\}]+{symm}{//}{TraditionalForm}}\)

\begin{equation}
\begin{array}{c}
\color{output}{symm}+6 \epsilon .\hat{\mathbf{x}} \epsilon' .\hat{\mathbf{x}} (i \xi \times k).\hat{\mathbf{x}}-6 \epsilon .\hat{\mathbf{y}} \epsilon' .\hat{\mathbf{y}}
(i \xi \times k).\hat{\mathbf{x}}-\\
\color{output}6 \epsilon .\hat{\mathbf{y}} \epsilon' .\hat{\mathbf{x}} (i \xi \times k).\hat{\mathbf{y}}-6 \epsilon .\hat{\mathbf{x}} \epsilon' .\hat{\mathbf{y}} (i \xi \times k).\hat{\mathbf{y}}
\end{array}
\end{equation}

\noindent where $\mathtt{symm}$ represents $[\epsilon \leftrightarrow k'] + [k \leftrightarrow k', \epsilon \leftrightarrow \epsilon']$].

After rearrangement according to the rules of the scalar triple product, this is equivalent to Equation~31 in \cite{mirone}.
This is the scattering factor contribution from the alternating term $\pm t_3(x^3-3 x y^2)$,
which is a dipole-quadrupole term.

\subsection{The physical meaning of the contraction results}

\paragraph{Extra peaks arising from asphericity of $T$}

The asphericity of the holmium atom results in extra peaks of Bragg order $2n \pm q$ (where $n \in \mathbb{N}$ and $q$ is the antiferromagnetic wavevector) appearing alongside the peaks of order $2n$ that one would normally expect to observe. Here we will examine the magnetic contribution to the scattering amplitude of the peak appearing at $2n+q$, for four polarisation channels ($\sigma \rightarrow \sigma$, $\pi \rightarrow \sigma$, $\sigma \rightarrow \pi$ and $\pi \rightarrow \pi$).

From \cite{mirone}, the contribution to the scattering amplitude at this peak, for a given field tensor $T$, is given by
\begin{equation}
F_{\epsilon,k \rightarrow \epsilon',k'} = \mathbb{C}\left(\epsilon' T \sum_{n=1} (i \xi \times)^n F^{\prime n}_1 \epsilon \right) + \mathbb{C}\left((k' \otimes \epsilon') T \sum_{n=1,3} (i \xi \times)^n F^{\prime n}_2 (k \otimes \epsilon) \right) / 2
\end{equation}
which originates from Equation~\ref{fcartesiancontractions}, where only the terms in odd $n$ are considered.

The $F$ prefactors are defined as follows:\\

\label{definingfprefactors}
\noindent\(\color{code}\mathtt{{F1' }_0=F_{10};}\)

\noindent\(\color{code}\mathtt{{F1' }_1=\frac{\left(F_{11}-F_{1-1}\right)}{2};}\)

\noindent\(\color{code}\mathtt{{F1' }_2=\frac{\left(2F_{10}-F_{11}-F_{1-1}\right)}{2};}\)

\noindent\(\color{code}\mathtt{{F2' }_0=F_{20};}\)

\noindent\(\color{code}\mathtt{{F2' }_1=\frac{\left(F_{2-2}-F_{22}+8F_{21}-8F_{2-1}\right)}{12};}\)

\noindent\(\color{code}\mathtt{{F2' }_2=\frac{\left(16F_{21}+16F_{2-1}-F_{2-2}-F_{22}-30F_{20}\right)}{24};}\)

\noindent\(\color{code}\mathtt{{F2' }_3=\frac{\left(F_{22}-F_{2-2}+2F_{2-1}-2F_{21}\right)}{12};}\)

\noindent\(\color{code}\mathtt{{F2' }_4=\frac{\left(6F_{20}-F_{22}-2F_{2-2}-4F_{21}-4F_{2-1}\right)}{24};}\)
\\

\noindent where the $F_{1,q}$ and $F_{2,q}$ originate from the treatment of the Holmium atom as the spherical case with a magnetic field perturbation, although we are not concerned with their definition here.

\paragraph{Redefining the scalar product}

We now want to create another definition of $\mathtt{scal}$, so that it decomposes $\mathtt{scal}[\vec{a},\vec{b}]$ into $a_x b_x + a_y b_y + a_z b_z$---this is necessary as we will be using the definitions from Equations~\ref{kdefinition} and \ref{epsilondefinition} to calculate the $\tau$ and $\theta$ dependence of the scattering amplitude.

First we define the scalar product function for various combinations of vectors, basis vectors and cross products with $\vec{\xi}$. We are making a new definition of $\mathtt{scal}$, and so we remove the previous one.\\

\noindent\(\color{code}\mathtt{{Clear}[{scal},\xi ];}\\
\mathtt{{Attributes}[{scal}]=\{{Orderless}\};}\)
\\

\noindent Now we define the scalar product of various expressions involving a cross-product with the magnetic basis vector $\vec{\xi}$, and any of the basis vectors $\vec{e}_1$, $\vec{e}_2$ or $\vec{e}_3$. In our co-ordinate system, $\vec{\xi} = (0,0,1) = \vec{e}_3$, so $\mathtt{scal}$ has been defined accordingly.
\\

\noindent\(\color{code}\mathtt{{scal}\left[i \xi \times (i \xi \times (i \xi \times {vector\_})),e_{{i\_}}\right]{:=}}\\
\mathtt{i \left\{-{vector}_y,{vector}_x,0\right\}[[i]];}\\
\mathtt{{scal}\left[i \xi \times (i \xi \times {vector\_}),e_{{i\_}}\right]{:=}\left\{{vector}_x,{vector}_y,0\right\}[[i]];}\\
\mathtt{{scal}\left[i \xi \times {vector\_},e_{{i\_}}\right]{:=}i \left\{-{vector}_y,{vector}_x,0\right\}[[i]];}\\
\mathtt{{scal}\left[{vector\_},e_{{i\_}}\right]{:=}{vector}_{\{x,y,z\}[[i]]};}\)
\\

Now we make a similar definition for scalar products with other vectors, such as $\vec{k}$ or $\vec{\epsilon}$.
\\

\noindent\(\color{code}\mathtt{{scal}[i \xi \times (i \xi \times (i \xi \times {vec1\_})),{vec2\_}]{:=}}\\
\mathtt{i \left(-{vec1}_y{vec2}_x+{vec1}_x{vec2}_y\right);}\\
\mathtt{{scal}[i \xi \times (i \xi \times {vec1\_}),{vec2\_}]{:=}\left({vec1}_x{vec2}_x+{vec1}_y{vec2}_y\right);}\\
\mathtt{{scal}[i \xi \times {vec1\_},{vec2\_}]{:=}i \left(-{vec1}_y{vec2}_x+{vec1}_x{vec2}_y\right);}\\
\mathtt{{scal}[{vec1\_},{vec2\_}]{:=}}\\
\mathtt{{Sum}\left[{vec1}_{\{x,y,z\}[[i]]}{vec2}_{\{x,y,z\}[[i]]},\{i,1,3\}\right];}\)

\paragraph{Considering the expected form of the solution}

Knowing that for the case of this particular peak the magnetic contribution to the scattering amplitude will have terms in $i(F_{1,1}-F_{1,-1})$, $i(F_{2,1}-F_{2,-1})$ and $i(F_{2,2}-F_{2,-2})$, we can initialise a small array, $\mathtt{magterm}$, to store these three terms.\\

\noindent\(\color{code}\mathtt{{magterm}=\{0,0,0\};}\)

\paragraph{Calculating the first part of the scattering amplitude expression}

Now we will try and find the first term, which will have a prefactor of $i(F_{1,1}-F_{1,-1})$). This term derives from the dipole-dipole contraction:\\

\noindent\(\color{code}\mathtt{{magterm}[[1]]=}\\
\mathtt{~~{Simplify}[}\\
\mathtt{~~{F1'}_1\mathbb{C}\left[\{\epsilon' \},\text{t2}\left(2z^2-x^2-y^2\right)+\text{t3}\left(x^3-3 x y^2\right),\{i \xi \times
\epsilon \}\right]\text{/.}}\\
\mathtt{~~\left.{t2}\to 1{/.}F_{11}-F_{1-1}\to \frac{1}{i}\right]}\)
\\

\noindent After simplification, the result is

\begin{equation}
\color{output}
\epsilon _y \epsilon' _x-\epsilon _x \epsilon' _y
\end{equation}
i.e. there is a contribution to the scattering amplitude of

\begin{equation}
i \left(F_{1,1}-F_{1,-1}\right)\left(\epsilon _y \epsilon' _x - \epsilon _x \epsilon' _y\right)
\end{equation}

This will be evaluated later on in more detail from the definitions of $\epsilon$, $\epsilon'$, $k$ and $k'$ in Equations~\ref{kdefinition} and \ref{epsilondefinition}, and re-expressed in terms of the incident angle $\theta$ and the antiferromagnetic parameters of the holmium system.

\paragraph{Calculating the second and third parts of the scattering amplitude expression}

We can now calculate the terms in $i(F_{2,1}-F_{2,-1})$ and $i(F_{2,2}-F_{2,-2})$. These originate from the contraction

\begin{equation}
\left( F^{\prime 1}_2 \mathbb{C}\left((\vec{k}'\otimes\vec{\epsilon}') T (i\vec{\xi}\times)(\vec{k}'\otimes\vec{\epsilon}')\right)
+ F^{\prime 3}_2 \mathbb{C}\left((\vec{k}'\otimes\vec{\epsilon}') T (i\vec{\xi}\times)^3 (\vec{k}'\otimes\vec{\epsilon}')\right) \right)/2
\end{equation}

where an operation of $i\vec{\xi}\times$ on a tensor is expanded binomially as a sum of operations on the vector components of the tensor. In Mathematica, we calculate\\

\noindent\(\color{code}\mathtt{{baseformagterms2and3}=}\\
\mathtt{~~{FullSimplify}[}\\
\mathtt{~~~~{F2' }_1\left(\mathbb{C}\left[\{{k' },\epsilon' \},{t2}\left(2z^2-x^2-y^2\right)+{t3}\left(x^3-3 x y^2\right),\right.\right.}\\
\mathtt{~~~~\{k,i \xi \times \epsilon \}]+}\\
\mathtt{~~~~\mathbb{C}\left[\{{k' },\epsilon' \},{t2}\left(2z^2-x^2-y^2\right)+{t3}\left(x^3-3 x y^2\right),\right.}\\
\mathtt{~~~~\{i \xi \times k,\epsilon \}])+}\\
\mathtt{~~~~{F2' }_3}\\
\mathtt{~~~~\left(\mathbb{C}\left[\{{k' },\epsilon' \},{t2}\left(2z^2-x^2-y^2\right)+{t3}\left(x^3-3 x y^2\right),\right.\right.}\\
\mathtt{~~~~\{k,i \xi \times (i \xi \times (i \xi \times \epsilon ))\}]+}\\
\mathtt{~~~~3\mathbb{C}\left[\{{k' },\epsilon' \},{t2}\left(2z^2-x^2-y^2\right)+{t3}\left(x^3-3 x y^2\right),\right.}\\
\mathtt{~~~~\{i \xi \times k,i \xi \times (i \xi \times \epsilon )\}]+}\\
\mathtt{~~~~3\mathbb{C}\left[\{{k' },\epsilon' \},{t2}\left(2z^2-x^2-y^2\right)+{t3}\left(x^3-3 x y^2\right),\right.}\\
\mathtt{~~~~\{i \xi \times (i \xi \times k),i \xi \times \epsilon \}]+}\\
\mathtt{~~~~\mathbb{C}\left[\{{k' },\epsilon' \},{t2}\left(2z^2-x^2-y^2\right)+{t3}\left(x^3-3 x y^2\right),\right.}\\
\mathtt{~~~~\{i \xi \times (i \xi \times (i \xi \times k)),\epsilon \}]){/.}{t2}\to 1]/2}\)\\

\noindent which gives the rather complicated result

\begin{equation}
\color{output}
\label{baseformagterms2and3}
\begin{array}{c}
\frac{1}{2} i \left(k_z \left({k' }_z \left(\epsilon _y \epsilon' _x-\epsilon _x \epsilon' _y\right)+\left(-{k'
}_y \epsilon _x+{k' }_x \epsilon _y\right) \epsilon' _z\right) \left(F_{2,-1}-F_{2,1}\right)+\right.\\
k_x \left({k' }_z \epsilon _z \epsilon' _y \left(-F_{2,-1}+F_{2,1}\right)+\right.\\
{k' }_y \left(\epsilon _z \epsilon' _z \left(-F_{2,-1}+F_{2,1}\right)+2 \left(\epsilon _x \epsilon' _x+\epsilon _y \epsilon
' _y\right) \left(F_{2,-2}-F_{2,2}\right)\right)+\\
\left.2 {k' }_x \left(\epsilon _y \epsilon' _x-\epsilon _x \epsilon' _y\right) \left(-F_{2,-2}+F_{2,2}\right)\right)+\\
k_y \left({k' }_z \epsilon _z \epsilon' _x \left(F_{2,-1}-F_{2,1}\right)+\right.\\
{k' }_x \left(\epsilon _z \epsilon' _z \left(F_{2,-1}-F_{2,1}\right)-2 \left(\epsilon _x \epsilon' _x+\epsilon _y \epsilon
' _y\right) \left(F_{2,-2}-F_{2,2}\right)\right)+\\
\left.\left.2 {k' }_y \left(\epsilon _y \epsilon' _x-\epsilon _x \epsilon' _y\right) \left(-F_{2,-2}+F_{2,2}\right)\right)\right)
\end{array}
\end{equation}

\noindent which can be used to calculate the second and third terms in $F$.

\paragraph{Calculating the second part of the scattering amplitude expression}

First we use Equation~\ref{baseformagterms2and3} to calculate the scattering amplitude term in $i(F_{2,1}-F_{2,-1})$. This involves taking only the coefficients of $iF_{2,1}$ (by symmetry, the $-iF_{2,-1}$ terms are identical), and discarding the other terms.\\

\noindent\(\color{code}\mathtt{{magterm}[[2]]=}\\
\mathtt{~~{FullSimplify}[}\\
\mathtt{~~~~\left.{baseformagterms2and3}{/.} F_{21}\to \frac{1}{i}{/.}F_{2-1}\to 0{/.} F_{22}\to 0{/.}F_{2-2}\to 0\right]}\)\\

\begin{equation}
\color{output}
\begin{array}{c}
\frac{1}{2} \left(k_z \left({k' }_z \left(-\epsilon _y \epsilon' _x+\epsilon _x \epsilon' _y\right)+\left({k'
}_y \epsilon _x-{k' }_x \epsilon _y\right) \epsilon' _z\right)+\right.\\
\left.\epsilon _z \left(-k_y \left({k' }_z \epsilon' _x+{k' }_x \epsilon' _z\right)+k_x \left({k'
}_z \epsilon' _y+{k' }_y \epsilon' _z\right)\right)\right)
\end{array}
\end{equation}

\paragraph{Calculating the third part of the scattering amplitude expression}

And now we find the term in $i(F_{2,2}-F_{2,-2})$, by the same method.\\

\noindent\(\color{code}\mathtt{{magterm}[[3]]=}\\
\mathtt{~~{FullSimplify}[}\\
\mathtt{~~~~\left.{baseformagterms2and3}{/.} F_{21}\to 0{/.}F_{2-1}\to 0{/.} F_{22}\to \frac{1}{i}{/.}F_{2-2}\to 0\right]}\)

\begin{equation}
\color{output}
\begin{array}{c}
k_y \left({k'}_y \left(\epsilon _y \epsilon' _x-\epsilon _x \epsilon' _y\right)+{k'}_x \left(\epsilon
_x \epsilon' _x+\epsilon _y \epsilon' _y\right)\right)-\\
k_x \left({k'}_x \left(-\epsilon _y \epsilon' _x+\epsilon _x \epsilon' _y\right)+{k'}_y \left(\epsilon _x \epsilon
' _x+\epsilon _y \epsilon' _y\right)\right)
\end{array}
\end{equation}

\paragraph{Reducing the scattering amplitude expression}

The three terms in $F$ that we have just calculated are:\\

\noindent\(\color{code}\mathtt{{magterm}{//}{TableForm}}\)

\begin{eqnarray}
\color{output}
& \epsilon _y \epsilon ' _x-\epsilon _x \epsilon ' _y
\\
\color{output}
& \begin{array}{l}
 \frac{1}{2} \left[k_z \left({k'}_z \left(-\epsilon _y \epsilon ' _x+\epsilon _x \epsilon ' _y\right)+\left({k'
}_y \epsilon _x-{k'}_x \epsilon _y\right) \epsilon ' _z\right) \right. \\
\left. +\epsilon _z \left(-k_y \left({k'}_z \epsilon '
_x+{k'}_x \epsilon ' _z\right)
+k_x \left({k'}_z \epsilon ' _y+{k'}_y \epsilon ' _z\right)\right)\right]
\end{array}
\\
& \begin{array}{l}
 k_y \left({k'}_y \left(\epsilon _y \epsilon ' _x-\epsilon _x \epsilon ' _y\right)+{k'}_x \left(\epsilon _x \epsilon
' _x+\epsilon _y \epsilon ' _y\right)\right)
\\
-k_x \left({k'}_x \left(-\epsilon _y \epsilon ' _x+\epsilon _x \epsilon '
_y\right) 
+{k'}_y \left(\epsilon _x \epsilon ' _x+\epsilon _y \epsilon ' _y\right)\right)
\end{array}
\end{eqnarray}

\noindent We introduce a shorthand notation and re-express the tensor products.\\

\noindent\(\color{code}\mathtt{i\left\{F_{11}-F_{1{-1}},F_{21}-F_{2{-1}},F_{22}-F_{2{-2}}\right\}}\\
\mathtt{~~\times {FullSimplify}[{magterm},}\\
\mathtt{~~\left\{\epsilon _xk_y+\epsilon _yk_x{==}{k\epsilon }_{xy},\epsilon _xk_z+\epsilon _zk_x{==}{k\epsilon }_{xz},\epsilon
_yk_z+\epsilon _zk_y{==}{k\epsilon }_{yz},\right.}\\
\mathtt{~~\epsilon' _x{k'}_y+\epsilon' _y{k'}_x{==}{k' \epsilon' }_{xy},\epsilon' _x{k'
}_z+\epsilon' _z{k'}_x{==}{k' \epsilon' }_{xz},}\\
\mathtt{~~\epsilon' _y{k'}_z+\epsilon' _z{k'}_y{==}{k' \epsilon' }_{yz},{k'}_x
\epsilon' _x-{k'}_y \epsilon' _y{==}{k' \epsilon' }_{{x2my2}},}\\
\mathtt{~~\left.\left.k_x \epsilon _x-k_y \epsilon _y{==}{k\epsilon }_{{x2my2}}\right\}\right]{//}{TableForm}{//}{TraditionalForm}}\)\\

\noindent So the reduced (simplified) form of the terms is

\begin{eqnarray}
\color{output}
& i \left(\epsilon _y \epsilon' _x-\epsilon _x \epsilon' _y\right) \left(F_{1,1}-F_{1,{-1}}\right) \\
& \frac{1}{2} i \left(F_{2,1}-F_{2,{-1}}\right) \left({k\epsilon }_{x,z} {k' \epsilon' }_{y,z}-{k\epsilon }_{y,z}
{k' \epsilon' }_{x,z}\right) \\
& i \left(F_{2,2}-F_{2,{-2}}\right) \left({k' \epsilon' }_{{x2my2}} {k\epsilon }_{x,y}-{k\epsilon }_{{x2my2}}
{k' \epsilon' }_{x,y}\right)
\end{eqnarray}

\paragraph{Finding the scattering amplitude matrices}

Now we will evaluate the expression for the scattering amplitude. Since $\epsilon$ and $\epsilon'$ can both be in either the $\pi$ or the $\sigma$ polarisation, we construct a $2\times 2$ matrix to represent the four combinations of $\epsilon \rightarrow \epsilon'$ in terms of $\theta$ and $\tau$.

First we evaluate the matrix
\(
F = \left(
\begin{array}{cc}
F_{\sigma' \leftarrow \sigma} & F_{\sigma' \leftarrow \pi} \\
F_{\pi' \leftarrow \sigma} & F_{\pi' \leftarrow \pi}
\end{array}
\right)
\)
or rather its three constituent terms which originate from the scattering amplitude expression calculated above.\\

\noindent\(\color{code}\mathtt{{Fmatrix}{:=}}\\
\mathtt{~~{Table}[}\\
\mathtt{~~~~{Simplify}[}\\
\mathtt{~~~~~~{magterm}[[i]]{/.}\left\{{vector\_}_{{xyz\_}}\to {vector}[\theta ,\tau ][[{xyz}]]\right\}{/.}}\\
\mathtt{~~~~~~~~\left.\left\{\epsilon \to \epsilon _{\pi \sigma },\epsilon' \to \epsilon' _{\pi \sigma' }\right\}{/.}\{x\to 1,y\to 2,z\to
3\}\right],\{i,1,3\},}\\
\mathtt{~~~~\{\pi \sigma' ,0,1\},\{\pi \sigma ,0,1\}];}\)
\\

\noindent The three terms are\\

\noindent\(\color{code}\mathtt{{Table}[{Fmatrix}[[i]]{//}{MatrixForm},\{i,1,3\}]{//}{TableForm}{//}}\\
\mathtt{~~{TraditionalForm}}\)

\begin{eqnarray}
\color{output}
\begin{array}{c}
 \left(
\begin{array}{ll}
 0 & -\mathrm{c} (\theta ) \mathrm{c} (\tau ) \\
 \mathrm{c} (\theta ) \mathrm{c} (\tau ) & -2 \mathrm{c} (\theta ) \mathrm{s} (\theta ) \mathrm{s} (\tau )
\end{array}
\right) \\
 \left(
\begin{array}{l}
 -\frac{1}{2} \mathrm{c} (2 \tau ) \mathrm{s} (2 \theta ) \mathrm{s} (\tau ) \\
\;\;\;\;\;\;
  \frac{1}{8} \mathrm{c} (\theta ) \mathrm{c} (\tau ) (-2 \mathrm{c} (2 \theta )+3 \mathrm{c} (2 (\theta -\tau
))-2 \mathrm{c} (2 \tau )+3 \mathrm{c} (2 (\theta +\tau ))+2) \\
 \frac{1}{8} \mathrm{c} (\theta ) \mathrm{c} (\tau ) (2 \mathrm{c} (2 \theta )-3 \mathrm{c} (2 (\theta -\tau ))+2 \mathrm{c} (2 \tau )-3 \mathrm{c} (2 (\theta +\tau ))-2) \\
\;\;\;\;\;\;
   \frac{1}{2}
\mathrm{c} ^2(\tau ) \mathrm{s} (4 \theta ) \mathrm{s} (\tau )
\end{array}
\right) \\
 \left(
\begin{array}{l}
 -2 \mathrm{c} (\theta ) \mathrm{c} ^2(\tau ) \mathrm{s} (\theta ) \mathrm{s} (\tau ) \\
\;\;\;\;\;\;
   \mathrm{c} (\theta ) \mathrm{c} (\tau ) \left(\mathrm{s} ^2(\theta ) \left(2 \mathrm{s} ^2(\tau )+1\right)-\mathrm{c}
^2(\theta ) \mathrm{s} ^2(\tau )\right) \\
 \mathrm{c} (\theta ) \mathrm{c} (\tau ) \left((\mathrm{c} (2 \tau )-2) \mathrm{s} ^2(\theta )+\mathrm{c} ^2(\theta ) \mathrm{s} ^2(\tau )\right) \\
\;\;\;\;\;\;
   \frac{1}{8} \mathrm{s} (4 \theta ) (\mathrm{s}
(3 \tau )-7 \mathrm{s} (\tau ))
\end{array}
\right)
\end{array}
\end{eqnarray}

\noindent where the last two $4\times4$ matrices have been condensed into a columnar form, and the abbreviations $\mathrm{s}$ and $\mathrm{c}$ stand for $\sin$ and $\cos$ respectively.

\paragraph{Fourier analysis of the $F$ matrix}

Since we are interested in the $+q$ satellite, we need to find the $e^{i\tau}$ Fourier component of the matrix $F$. The following code breaks $F$ into $e^{i n\tau}$ components and extracts the first Fourier coefficients.\\

\noindent\(\color{code}\mathtt{{Fourier1}=}\\
\mathtt{{Table}[}\\
\mathtt{~~{TrigReduce}[}\\
\mathtt{~~~~{PadRight}[{CoefficientList}[}\\
\mathtt{~~~~~~{Expand}[{TrigExpand}[{Fmatrix}[[i,{var1},{var2}]]]{/.}}\\
\mathtt{~~~~~~~~{Sin}[{n\_.} \tau ]\to \frac{1}{2i}\left(e^{i n \tau }-e^{-i n \tau }\right){/.}}\\
\mathtt{~~~~~~~~\left.\left.\left.{Cos}[{n_.} \tau ]\to \frac{1}{2}\left(e^{i n \tau }+e^{-i n \tau }\right)\right]{/.}e^{i \tau }\to {TERM},{TERM}\right],2\right][[}\\
\mathtt{~~~~~~~~2]]],\{i,1,3\},\{{var1},1,2\},\{{var2},1,2\}];}\)
\\

\noindent So the coefficient of the scattering amplitude matrix multiplying the $e^{i\tau}$ Fourier component is:\\

\noindent\(\color{code}\mathtt{{Do}\left[{Print}\left[{\mathtt{"}+\mathtt{"}},i\left\{F_{11}-F_{1-1},F_{21}-F_{2-1},F_{22}-F_{2-2}\right\}[[i]]{//}\right.\right.}\\
\mathtt{~~{TraditionalForm},{Fourier1}[[i]]{//}{MatrixForm}{//}}\\
\mathtt{~~{TraditionalForm}],\{i,1,3\}]}\)

\begin{equation}
\color{output}
\begin{array}{c}
i \left(F_{1,1}-F_{1,{-1}}\right)\left(
\begin{array}{cc}
 0 & -\frac{\mathrm{c} (\theta )}{2} \\
 \frac{\mathrm{c} (\theta )}{2} & \frac{1}{2} i \mathrm{s} (2 \theta )
\end{array}
\right)
\\
+i \left(F_{2,1}-F_{2,{-1}}\right)
\left(
\begin{array}{cc}
 -\frac{1}{8} i \mathrm{s} (2 \theta ) & \frac{1}{32} (3 \mathrm{c} (\theta )+\mathrm{c} (3 \theta )) \\
 \frac{1}{32} (-3 \mathrm{c} (\theta )-\mathrm{c} (3 \theta )) & -\frac{1}{16} i \mathrm{s} (4 \theta )
\end{array}
\right)
\\
+i \left(F_{2,2}-F_{2,{-2}}\right)
\left(
\begin{array}{cc}
 \frac{1}{8} i \mathrm{s} (2 \theta ) & \frac{1}{32} (3 \mathrm{c} (\theta )-7 \mathrm{c} (3 \theta )) \\
 \frac{1}{32} (7 \mathrm{c} (3 \theta )-3 \mathrm{c} (\theta )) & \frac{7}{16} i \mathrm{s} (4 \theta )
\end{array}
\right)
\end{array}
\end{equation}

The above expression is the scattering amplitude matrix for the $n+q$ satellite diffraction peak of holmium\footnote{This can be shown by considering components of a wave incident at angle $\theta$ being reflected off adjacent crystal layers $z_1$, $z_2$ with scattering coefficients $e^{i\tau_1}$ and $e^{i\tau_2}$. The electric fields of the two reflected waves will be given by $e^{i\tau_1} e^{ikx}$ and $e^{i\tau_2} e^{ik(x+2d \sin \theta)}$. The condition for these to be in phase leads to $2d \sin \theta = q \lambda$, a modified Bragg diffraction condition.}. This can be used to obtain theoretical predictions for the amplitudes of the various scattering peaks for photons of different energies---which can then be compared with measurements made at the ESRF.

\section{Conclusions}

We have taken the contraction method established in \cite{mirone} and implemented it as an efficient computer algorithm in a Mathematica program. The program was designed to take a general (polynomial) field tensor and contract it with various combinations of $\vec{k}$ and $\vec{\epsilon}$ with the magnetic vector $\vec{\xi}$ acting as a rotation derivative. The program output scattering amplitude expressions in terms of components of $\vec{k}$ and $\vec{\epsilon}$.

An example of the program operation was presented here for the case of spiral antiferromagnetic holmium, where the vector $\vec{\xi}$ rotates round through the crystal layering structure with an antiferromagnetic wavevector $q$. Using the non-spherical field tensor for this material, the dipole-dipole, quadrupole-quadrupole and mixed dipole-quadrupole corrections to the scattering factors which were obtained for the case of a spherical atom\cite{hm} were calculated.

The contraction method was then used to find the amplitude of the diffraction peak with Bragg number $2n \pm q$, for the $\sigma' \leftarrow \sigma$, $\sigma' \leftarrow \pi$, $\pi' \leftarrow \sigma$ and $\pi' \leftarrow \pi$ polarisations, via a Fourier analysis of the scattering amplitude function. These scattering amplitudes could then be used to predict the intensities of the satellite peaks observed in RMXS experiments on Holmium.

\paragraph{Further work}
This program can be developed further to allow for more complex forms of field tensors, such as those involving cross products with the photon tensors (see Appendix~\ref{crossproducts}). The reliability of the contraction method\cite{mirone} in predicting scattering peaks for other materials is yet to be tested.

The extension of the program to cope with contractions of more than three tensors is relatively trivial, involving only the addition of a few more rules to reduce the number of tensors in the same way that the present program reduces it from three to two.

A more complicated problem would be to contract a set of tensors with restrictions on the possible contractions, such as in Equation~25 of \cite{mirone}.

\section{Acknowledgements}

I am grateful to Alessandro Mirone for his advice on the programming and for his patience in explaining the theoretical details of his paper\cite{mirone}.

\appendix

\section{Extension of the program}
\label{crossproducts}

%Based on \texttt{ advanced1.nb}, but edited to look better when it gets LaTeXified

\subsection{Tensor contraction}

\paragraph{Two tensors}
First we define the contraction sum for two tensors of any rank.\\

\noindent\(\color{code}\mathtt{{Attributes}[{Con}]=\{{Orderless}\};}\\
\mathtt{{Attributes}[T] = \{{Orderless}\};}\\
\mathtt{{Con}[{x\_\_},0]{:=}0;}\\
~\\
\mathtt{{Con}[T[{a\_}, {al\_\_\_}],T[{bl\_\_\_}]]{:=}{Module}[\{{res}\},{res}=0;}\\
\mathtt{n={Length}[{List}[{bl}]];}\\
\mathtt{{Do}[}\\
\mathtt{ {res}={res} + {scal}\left[a{/.}\left\{x\to e_1,y\to e_2,z\to e_3\right\},{List}[{bl}][[i]]{/.}\left\{x\to
e_1,y\to e_2,z\to e_3\right\}\right] }\\
\mathtt{{ConTmp}[ T[{al}], {Delete}[ T[{bl}],i] ]}\\
\mathtt{,\{i,1,n\}}\\
\mathtt{];}\\
\mathtt{{res}}\\
\mathtt{];}\\
\mathtt{{Con}[T[{c\_\_\_}]]{:=}0;}\\
\mathtt{{Con}[1,{t\_\_\_}]{:=}{Con}[t];}\\
\mathtt{{Con}[]{:=}1;}\\
\mathtt{{  }T[]=1; }\\
\mathtt{T[\{\}]=1;}\\
\mathtt{T[\{{x\_\_}\}]=T[x];}\)
\\

\paragraph{Three tensors}
Now we define the contraction sum for three tensors (by first eliminating one of the three, and then reverting to the two-tensor contraction).
\\

\noindent\(\color{code}\mathtt{{Con}[T[{al\_\_\_}],T[{bl\_\_\_}],T[{cl\_\_\_}]]{:=}{Module}[\{{res},m,n\},{res}=0;}\\
\mathtt{m={Length}[T[{bl}]];}\\
\mathtt{n={Length}[T[{cl}]];}\\
\mathtt{{Do}[}\\
\mathtt{ {res}=}\\
\mathtt{{res} + {scal}\left[{List}[{al}][[1]]{/.}\left\{x\to e_1,y\to e_2,z\to e_3\right\},\right.}\\
\mathtt{\left.{List}[{bl}][[i]]{/.}\left\{x\to e_1,y\to e_2,z\to e_3\right\}\right] }\\
\mathtt{{ConTmp}[ {Delete}[ T[{al}],1], {Delete}[ T[{bl}],i],T[{cl}] ]}\\
\mathtt{,\{i,1,m\}}\\
\mathtt{];}\\
\mathtt{{Do}[}\\
\mathtt{ {res}=}\\
\mathtt{{res} + {scal}\left[{List}[{al}][[1]]{/.}\left\{x\to e_1,y\to e_2,z\to e_3\right\},\right.}\\
\mathtt{\left.{List}[{cl}][[i]]{/.}\left\{x\to e_1,y\to e_2,z\to e_3\right\}\right] }\\
\mathtt{{ConTmp}[ {Delete}[ T[{al}],1],T[{bl}],{Delete}[ T[{cl}],i] ]}\\
\mathtt{,\{i,1,n\}}\\
\mathtt{];}\\
\mathtt{{res}}\\
\mathtt{];}\)
\\

As a temporary measure (for debugging), we make all \texttt{ConTmp} be executed as \texttt{Con}.
\\

\noindent\(\color{code}\mathtt{{ConTmp}={Con};}\)

\subsection{Contracting polynomials with cross products}

We will construct several functions, all defined by rules, along which the contraction is passed. Each function executes its own simplifications
and contractions, and then invokes the next one in the sequence.

\paragraph{Expanding brackets}
The function $\mathbb{C}$ rewrites the polynomial in expanded form, and passes the contraction to $\mathbb{C}\mathtt{2}$, the second contracting function.
\\

\noindent\(\color{code}\mathtt{\mathbb{C}[{al\_},{anything\_},{cl\_}]{:=}{\mathbb{C}2}[{al},{Expand}[{anything}],{cl}];}\)

\paragraph{Expanding contractions of sums}
If we are a contracting a sum of terms, then $\mathbb{C}\mathtt{3}$ splits it into a sum of contractions...
\\

\noindent\(\color{code}\mathtt{{\mathbb{C}2}[{al\_},{Plus}[{thing1\_},{morethings\_\_}],{cl\_}]{:=}}\\
\mathtt{{Sum}[{\mathbb{C}2}[{al},{Join}[{List}[{thing1}],{List}[{morethings}]][[i]],{cl}],}\\
\mathtt{\{i,1,{Length}[{Join}[{List}[{thing1}],{List}[{morethings}]]]\}];}\)
\\

...until we we have nothing left to expand into sums: then it passes the contraction on to $\mathbb{C}\mathtt{3}$, the third contraction function
\\

\noindent\(\color{code}\mathtt{{\mathbb{C}2}[{al\_},{Sum}[{onething\_}],{cl\_}]{:=}{\mathbb{C}3}[{al},{onething},{cl}];}\)

\paragraph{Taking sums outside $\times$ operator}
Now $\mathbb{C}\mathtt{3}$ expands the sums within cross products into sums of contractions

\noindent\(\color{code}\mathtt{{\mathbb{C}3}[{al\_},}\\
\mathtt{{Times}[{prefactors\_\_\_},{vec1\_}\times {Plus}[{vec2a\_},{vec2b\_\_}]],{cl\_}]{:=}}\\
\mathtt{{Times}{@@}{Complement}\left[\{{prefactors}\},\left\{x,y,z,\mathbf{x},\mathbf{y},\mathbf{z},x^{\_},y^{\_},z^{\_}\right\}\right]}\\
\mathtt{{Times}[}\\
\mathtt{{Sum}\left[{\mathbb{C}3}\left[{al},{Times}{@@}{Intersection}\left[\{{prefactors}\},\left\{x,y,z,\mathbf{x},\mathbf{y},\mathbf{z},x^{\_},y^{\_},z^{\_}\right\}\right]\right.\right.}\\
\mathtt{{vec1}\times ({vec2a}+{vec2b})[[i]],{cl}],\{i,1,{Length}[{vec2a}+{vec2b}]\}]];}\)

\noindent\(\color{code}\mathtt{{\mathbb{C}3}[{al\_},}\\
\mathtt{{Times}[{prefactors\_\_\_},{Plus}[{vec1a\_},{vec1b\_\_}]\times {vec2\_}],{cl\_}]{:=}}\\
\mathtt{{Times}{@@}{Complement}\left[\{{prefactors}\},\left\{x,y,z,\mathbf{x},\mathbf{y},\mathbf{z},x^{\_},y^{\_},z^{\_}\right\}\right]}\\
\mathtt{{Times}[}\\
\mathtt{{Sum}\left[{\mathbb{C}3}\left[{al},{Times}{@@}{Intersection}\left[\{{prefactors}\},\left\{x,y,z,\mathbf{x},\mathbf{y},\mathbf{z},x^{\_},y^{\_},z^{\_}\right\}\right]\right.\right.}\\
\mathtt{({vec1a}+{vec1b})[[i]]\times {vec2},{cl}],\{i,1,{Length}[{vec1a}+{vec1b}]\}]];}\)

\paragraph{Taking powers outside $\times$ operator}
Now $\mathbb{C}\mathtt{3}$ can also expand powers within cross products, and re-write the expression as a sum of cross products with each of the variable
on the right of the cross-product operator.
\\

\noindent\(\color{code}\mathtt{{\mathbb{C}3}[{al\_},{Times}[{prefactors\_\_\_},{Cross}[{a\_},{Power}[{xyz1\_},{power\_}]]],{cl\_}]{:=}}\\
\mathtt{{Module}[\{i,{xyz}\},{xyz}={Table}[{xyz1},\{i,1,{power}\}];}\\
\mathtt{{Times}{@@}{Complement}\left[\{{prefactors}\},\left\{x,y,z,\mathbf{x},\mathbf{y},\mathbf{z},x^{\_},y^{\_},z^{\_}\right\}\right]}\\
\mathtt{{Sum}[}\\
\mathtt{{\mathbb{C}3}[{al},}\\
\mathtt{{Times}[{Times}{@@}{Intersection}[\{{prefactors}\},}\\
\mathtt{\left.\left.\left.\left\{x,y,z,\mathbf{x},\mathbf{y},\mathbf{z},x^{\_},y^{\_},z^{\_}\right\}\right],{Times}{@@}{Delete}[{xyz},i],{cross}[a,{xyz}[[i]]]\right],{cl}\right],}\\
\mathtt{\{i,1,{Length}[{xyz}]\}]}\\
\mathtt{]}\)

\paragraph{Taking products outside $\times$ operator}
$\mathbb{C}\mathtt{3}$ also expands products within cross products, similarly re-expressing the polynomial as a sum of cross product operations on each of the variables on the right of the $\times$ sign.
\\

\noindent\(\color{code}\mathtt{{\mathbb{C}3}[{al\_},{Times}[{prefactors\_\_\_},{Cross}[{a\_},{Times}[{xyz1\_},{xyz2\_\_}]]],{cl\_}]{:=}}\\
\mathtt{{Module}[\{i,{xyz}\},}\\
\mathtt{{xyz}={Flatten}[{Join}[{List}[{xyz1}],{List}[{xyz2}]]{/.}}\\
\mathtt{\left.\left\{{anything\_}^{{power\_}}\to {Table}[{anything},\{i,1,{power}\}]\right\}\right];}\\
\mathtt{{Times}{@@}{Complement}\left[\{{prefactors}\},\left\{x,y,z,\mathbf{x},\mathbf{y},\mathbf{z},x^{\_},y^{\_},z^{\_}\right\}\right]}\\
\mathtt{{Sum}[}\\
\mathtt{{\mathbb{C}3}[{al},}\\
\mathtt{{Times}[{Times}{@@}{Intersection}[\{{prefactors}\},}\\
\mathtt{\left.\left.\left.\left\{x,y,z,\mathbf{x},\mathbf{y},\mathbf{z},x^{\_},y^{\_},z^{\_}\right\}\right],{Times}{@@}{Delete}[{xyz},i],{cross}[a,{xyz}[[i]]]\right],{cl}\right],}\\
\mathtt{\{i,1,{Length}[{xyz}]\}]}\\
\mathtt{]}\)
\\

\paragraph{Introducing the \texttt{cross} operator}
We now re-express any simple cross product within a contraction as a function of the cross product operator.
\\

\noindent\(\color{code}\mathtt{{\mathbb{C}3}[{al\_},{Times}[{prefactors\_\_\_},{Cross}[{a\_},{b\_}]],{cl\_}]{:=}}\\
\mathtt{{Times}{@@}{Complement}\left[\{{prefactors}\},\left\{x,y,z,\mathbf{x},\mathbf{y},\mathbf{z},x^{\_},y^{\_},z^{\_}\right\}\right]}\\
\mathtt{{\mathbb{C}3}\left[{al},{Times}{@@}{Intersection}\left[\{{prefactors}\},\left\{x,y,z,\mathbf{x},\mathbf{y},\mathbf{z},x^{\_},y^{\_},z^{\_}\right\}\right]{cross}[a,b],\right.}\\
\mathtt{{cl}]}\)

\paragraph{Defining the cross product operator}
~\\

\noindent\(\color{code}\mathtt{{cross}[{a\_},{b\_}]{:=}}\\
\mathtt{{Cross}[(a{/.}\{x\to \mathbf{x},y\to \mathbf{y},z\to \mathbf{z}\}{/.}\{\mathbf{x}\to \{1,0,0\},\mathbf{y}\to \{0,1,0\},\mathbf{z}\to
\{0,0,1\}\}),}\\
\mathtt{(b{/.}\{x\to \mathbf{x},y\to \mathbf{y},z\to \mathbf{z}\}{/.}\{\mathbf{x}\to \{1,0,0\},\mathbf{y}\to \{0,1,0\},\mathbf{z}\to \{0,0,1\}\})].\{x,y,z\}}\)

\paragraph{Correcting a small glitch}

We will get some dodgy terms arising from the program attempting to also contract minus signs as separate entities. We can patch over this with the
following rule:
\\

\noindent\(\color{code}\mathtt{{cross}[\_,-1]{:=}0}\)

\subsection{Contracting polynomials now that the cross-products have been removed}

Now the cross products in contractions have been all re-expressed as simple sums of scalar products, we can continue with the contractions as normal.

\paragraph{Taking summations outside}
The result from all the above simplifications might still be expandable: so any summations must therefore be taken outside the contraction sign again.
\\

\noindent\(\color{code}\mathtt{{\mathbb{C}3}[\{{al\_\_}\},{Plus}[{x\_},{y\_\_}],\{{cl\_\_}\}]{:=}}\\
\mathtt{{Module}[\{i\},{xy}={Expand}[x+y];{Sum}[{\mathbb{C}3}[\{{al}\},{xy}[[i]],\{{cl}\}],\{i,1,{Length}[{xy}]\}]]}\)

\paragraph{Executing the contractions}
Now we need to find an efficient way to take each term and extract the $xyz$-dependent components.

First we separate out the terms containing $x y z$ from the rest---this is so constant factors \textit{ etc.} don't get accidentally contracted when
they shouldn't.
\\

\noindent\(\color{code}\mathtt{{\mathbb{C}3}[\{{al\_\_}\},}\\
\mathtt{{Times}[{prefactor\_}{/;}{FreeQ}[{FullForm}[{prefactor}],x]{/;}{FreeQ}[{FullForm}[{prefactor}],y]{/;}}\\
\mathtt{{FreeQ}[{FullForm}[{prefactor}],z],{xyz\_\_}],\{{cl\_\_}\}]{:=}}\\
\mathtt{{Times}[{prefactor}] {\mathbb{C}3}[\{{al}\},{Times}[{xyz}],\{{cl}\}];}\)
\\

Now we convert the polynomial (which is now in $x y z$ form) into a tensorial form that can be used directly by the function \texttt{Con}.
\\

\noindent\(\color{code}\mathtt{{\mathbb{C}3}[\{{al\_\_}\},{Times}[{xyz1\_},{xyz2\_\_}],\{{cl\_\_}\}]{:=} {Module}[\{i,{power},n,{xyz}\},}\\
\mathtt{{xyz}{:=}{List}[{xyz1},{xyz2}]{/.}{x\_}^{{power\_}}{->}{Table}[x,\{i,1,{power}\}];}\\
\mathtt{n={Length}[{xyz}];}\\
\mathtt{{Con}[T[{al}],T[{Flatten}[{Table}[{xyz}[[i]],\{i,1,n\}]]],T[{cl}]]}\\
\mathtt{]}\)

\noindent\(\color{code}\mathtt{{\mathbb{C}3}[\{{al\_\_}\},{Power}[{xyz\_},{power\_}],\{{cl\_\_}\}]{:=} {Module}[\{i,n\},}\\
\mathtt{{Con}[T[{al}],T[{Table}[{xyz},\{i,1,{power}\}]],T[{cl}]]}\\
\mathtt{]}\)

\noindent\(\color{code}\mathtt{{\mathbb{C}3}[{al\_},0,{cl\_}]{:=}0;}\)

\subsection{Defining the scalar product symbolically}

\noindent\(\color{code}\mathtt{{Clear}[{scal},\xi ];}\\
\mathtt{{Attributes}[{scal}]=\{{Orderless}\};}\\
\mathtt{{scal}\left[{vector\_},e_{{i\_}}\right]{:=}{vector}.\{\mathbf{x},\mathbf{y},\mathbf{z}\}[[i]];}\\
\mathtt{{scal}[{vec1\_},{vec2\_}]{:=}{vec1}.{vec2};}\)

\subsection{Application: chiral system}

The contraction mechanism has been defined to work on any kind of polynomial tensor involving a cross-product with another vector. Now we test it
on the case of a chiral system, which has the field tensor

\begin{equation}
T = (1+z \alpha  \mathbf{z}\times )\left(x^2-y^2\right)
\end{equation}

\paragraph{Dipole-dipole correction}
We can see that this result is what you would expect from a contraction of \(x^2-y^2\) with $\epsilon ' $ and
$\epsilon $---the cross-product term in $\alpha $ has no effect as it has rank 3:
\\

\noindent\(\color{code}\mathtt{\mathbb{C}\left[\{\epsilon ' \},\left(x^2-y^2\right)+\alpha  z \mathbf{z}\times \left(x^2-y^2\right),\{\epsilon \}\right]}\)

\begin{equation}
\color{output}
2 \epsilon .\mathbf{x} \epsilon ' .\mathbf{x}-2 \epsilon .\mathbf{y} \epsilon ' .\mathbf{y}
\end{equation}

\paragraph{Mixed (dipole-quadrupole) correction}
Here only the term in $\alpha $ affects the result.
\\

\noindent\(\color{code}\mathtt{\mathbb{C}\left[\{\epsilon ' \},\left(x^2-y^2\right)+\alpha  z \mathbf{z}\times \left(x^2-y^2\right),\{k,\epsilon \}\right]}\)

\begin{equation}
\color{output}
\begin{array}{c}
\alpha  \left(4 (k.\mathbf{z} \epsilon .\mathbf{y}+k.\mathbf{y} \epsilon .\mathbf{z}) \epsilon ' .\mathbf{x}+4 (k.\mathbf{z} \epsilon
.\mathbf{x}+k.\mathbf{x} \epsilon .\mathbf{z}) \epsilon ' .\mathbf{y} \right.
\\
\left. +4 (k.\mathbf{y} \epsilon .\mathbf{x}+k.\mathbf{x} \epsilon .\mathbf{y})
\epsilon ' .\mathbf{z}\right)
\end{array}
\end{equation}

\paragraph{Quadrupole-quadrupole correction}
Now we are looking at the whole expression for the field tensor
\\

\noindent\(\color{code}\mathtt{\mathbb{C}\left[\{{k' },\epsilon ' \},\left(x^2-y^2\right)+\alpha  z \mathbf{z}\times \left(x^2-y^2\right),\{k,\epsilon
\}\right]}\)

\begin{equation}
\color{output}
\begin{array}{c}
2 k.\epsilon '  {k' }.\mathbf{x} \epsilon .\mathbf{x}-2 k.\epsilon '  {k' }.\mathbf{y} \epsilon .\mathbf{y}+2
k.{k' } \epsilon .\mathbf{x} \epsilon ' .\mathbf{x}\\
+ 2 k.\mathbf{x} ({k' }.\mathbf{x} \epsilon .\epsilon ' +{k' }.\epsilon  \epsilon ' .\mathbf{x})-2 k.{k'
} \epsilon .\mathbf{y} \epsilon ' .\mathbf{y}-2 k.\mathbf{y} ({k' }.\mathbf{y} \epsilon .\epsilon ' +{k' }.\epsilon
 \epsilon ' .\mathbf{y})
\end{array}
 \end{equation}

\end{document}